\documentclass[baaa]{baaa}

\usepackage[pdftex]{hyperref}
\usepackage{subfigure}
\usepackage{natbib}
\usepackage{helvet,soul}
\usepackage[font=small]{caption}

\contriblanguage{1}

\contribtype{1}

\thematicarea{2}

\title{Gas accretion onto the disc of a \\ simulated Milky Way-mass galaxy }

\titlerunning{Gas accretion onto a simulated galactic disc}

\author{
F.G. Iza\inst{1,2},
S.E. Nuza\inst{1,2}
\&
C. Scannapieco\inst{2}
}

\authorrunning{Iza et al.}

\contact{fiza@iafe.uba.ar}

\institute{Instituto de Astronomía y Física del Espacio, CONICET--UBA, Argentina
\and
Departamento de Física, Facultad de Ciencias Exactas y Naturales, UBA, Argentina
}

\resumen{
En el paradigma estándar de formación y evolución de galaxias, la componente bariónica de las galaxias se forma a partir del colapso y la condensación de gas dentro de halos de materia oscura, y posteriormente crece como resultado de la acreción continua de material gaseoso, tanto en forma difusa como en colisiones con otros sistemas. Luego de un período inicial donde el crecimiento del halo es rápido y violento, el gas se establece en una estructura soportada por rotación a partir de la cual se forma, posteriormente, el disco estelar. Las estrellas evolucionan y retornan  gas químicamente enriquecido al medio interestelar, principalmente a través de explosiones de supernova tipo II. En la región del disco, la acreción cosmológica de gas junto con los flujos salientes producidos por las supernovas afectan las propiedades hidrodinámicas y estructurales del disco, produciendo flujos de gas tanto verticales como radiales. En este trabajo, utilizamos una simulación del proyecto Auriga, un conjunto de simulaciones cosmológicas magnetohidrodinámicas del tipo {\it zoom-in}
para estudiar las dependencias temporal y radial de la acreción de gas hacia el disco galáctico en galaxias espirales. Investigamos también la evolución del disco, haciendo énfasis en el escenario {\it inside-out}, una de las hipótesis fundamentales de los modelos de evolución química de la Galaxia.
}

\abstract{
In the standard paradigm of galaxy formation and evolution, the baryonic component of galaxies forms from the collapse and condensation of gas within dark matter haloes, and later grows from continuous accretion of gaseous mass, both in diffuse form and in mergers with other systems. After a first period of rapid and violent halo growth, the gas settles into a rotationally-supported structure, eventually giving rise to the formation of a stellar disc. Stars evolve and return chemically-processed gas and energy to the interstellar medium, mainly through Type II supernova explosions. In the disc region, the cosmological accretion of gas combines with the outflows resulting from supernovae, affecting the hydrodynamical and structural properties of the disc and producing gas flows in the vertical and radial directions. In this work, we use a simulation of the Auriga Project, a suite of magneto-hydrodynamical, zoom-in cosmological simulations of Milky Way-like galaxies, to study the temporal and radial dependencies of gas accretion onto the disc. We also investigate the disc evolution, focusing on the {\it inside-out} disc formation scenario, which is one of the fundamental hypotheses of chemical evolution models of the Galaxy.
}


\begin{document}

\maketitle

\section{Introduction} \label{sec:intro}

The formation of structures in the Universe 
results from the amplification of small fluctuations in the primordial density field. This gives rise to a hierarchical clustering process in which 
mass accretion and mergers with smaller substructures drive the growth of galactic haloes.
In particular, the accretion of gas  
 is responsible of providing the fuel needed to build up the luminous component of galaxies
 and induces gas flows \citep{Bilitewski2012} that might contribute to settle a chemical gradient in the disc. Moreover, gas accretion plays a key role in the fate of galaxies and their morphological evolution \citep{Scannapieco2015}.

On the other hand, chemical evolution models (CEMs) and cosmological simulations show that, in order to reproduce the properties of  Milky Way-like galaxies, a sustained accretion of gas during --at least-- the last 5 to 7 Gyr of galactic evolution is necessary \citep{Kubryk2015, Nuza2014, Nuza2019}. CEMs also show that the Galactic disc might preferentially form from the “inside-out”, with the number of newly formed stars per unit time decreasing towards the disc outskirts \citep{Larson1976, Chiappini2001}.

Owing to the complexity of the physics involved in the process of galaxy formation and evolution, much recent effort has been devoted to produce high-resolution simulations which aim at describing the evolution of Milky Way-mass galaxies. Given that galaxies of similar mass show vastly different properties, it is also important to relate particular evolution histories to gas accretion processes and late-time morphology.

In this work, we compute gas flows onto the disc of a simulated galaxy from the Auriga Project \citep{Grand2017}, a set of zoom-in simulations performed using the magnetohydrodynamics (MHD)  cosmological code {\sc arepo} \citep{Springel2010}. 
We study the temporal dependency of the  total accretion onto the disc, calculating inflow, outflow and net accretion rates. In order to relate our analysis to the inside-out formation scenario, we also compute the radial dependency of gas accretion for the simulated galaxy.

This proceeding is organised as follows. In Sec.~2 we present the main characteristics of the {\sc arepo} code and give a brief description of the properties of the simulated galaxy at redshift zero, as well as the parameters adopted to follow the cosmological growth of the disc. In Sec.~3 we show the main results of this work: gas accretion as a function of time and radius. In Sec.~4 we summarise our findings.

\section{The simulated galaxy} \label{sec:sims}

In this work, we analyse a simulated galaxy --labelled Au6-- from the Auriga Project \citep{Grand2017}, a suite of 30 haloes simulated at high resolution using {\scshape arepo}, an $N$-body MHD  cosmological code \citep{Springel2010}.
The Au6 galaxy has a well-formed galactic disc at $z=0$ and its main structural properties are summarised as follows: $R_{200} = 213.83~\mathrm{kpc}$ (halo virial radius), $M_{200} = 104.39 \times 10^{10}~\mathrm{M}_\odot$ (halo virial mass), and $R_\mathrm{opt} = 26~\mathrm{kpc}$ (optical radius).

\begin{figure}[!t]
    \centering
    \includegraphics[width=0.8\columnwidth]{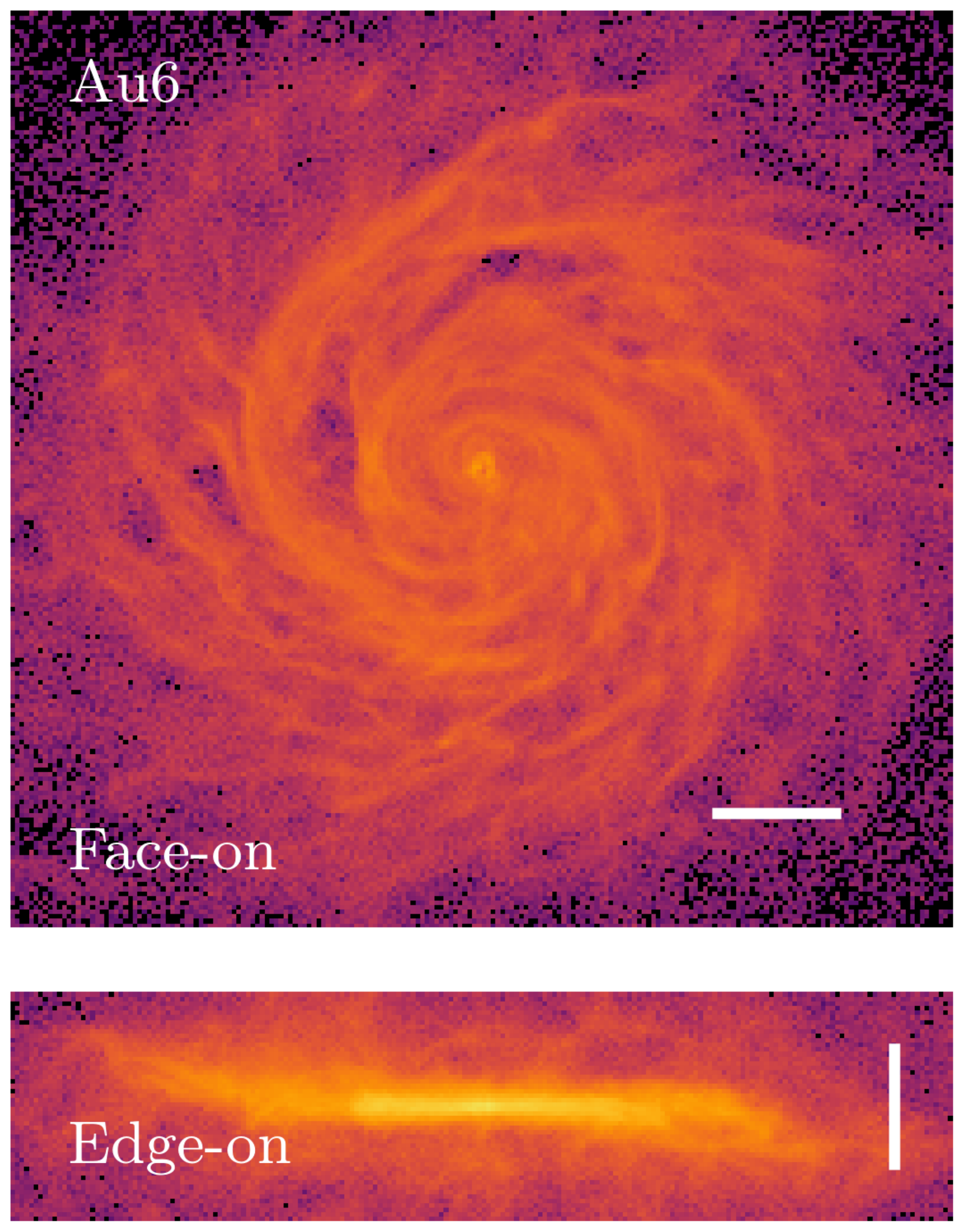}
    \caption{Gas surface density for the  Au6 simulated galaxy at redshift $z=0$. To guide the eye, white bars indicate a scale of $10~\mathrm{kpc}$. The colour map spans five orders of magnitude in projected density. \emph{Top panel:} face-on view, $xy$-plane. \emph{Bottom panel:} edge-on view, $xz$-plane.}
    \label{fig:dmaps}
\end{figure}


We use a simple definition of the disc region at each time, using  two parameters: a maximum disc radius $R_\mathrm{d}$ and a disc height $h_\mathrm{d}$ (with this definition, the disc extends from $-h_\mathrm{d}/2$ to $+h_\mathrm{d}/2$ in the $z$-direction). Fiducial values are $R_\mathrm{d}=30~\mathrm{ckpc}$ and $h_\mathrm{d}=2~\mathrm{ckpc}$, where ``c'' stands for comoving. Note that we use comoving coordinates to better follow the growth of the galaxy and, in particular, the disc. In order to perform our analysis related to the radial dependence of accretion, we divide the disc into 30 radial, $1~\mathrm{ckpc}$-wide bins.
Face-on and edge-on gaseous surface density maps for Au6 at $z=0$ are shown in Fig.~\ref{fig:dmaps}. 

\begin{figure}[!t]
    \centering
    \includegraphics[width=0.9\columnwidth]{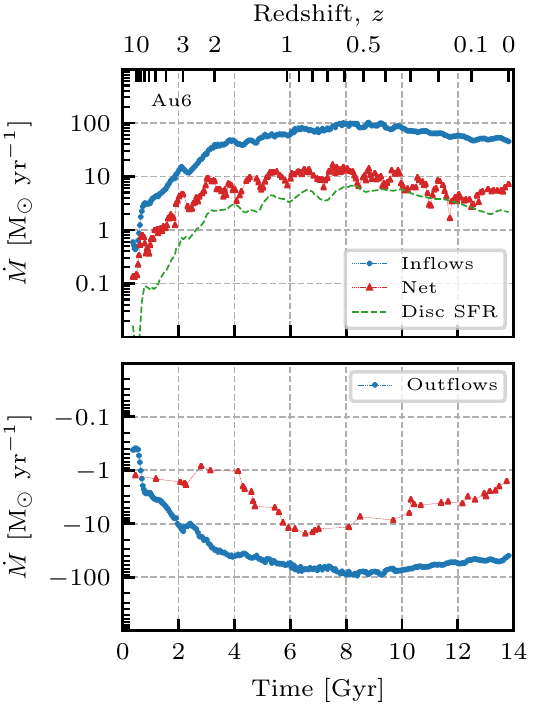}
    \caption{Accretion rate of gas as a function of time for the Au6 simulated galaxy. In both panels the red curves indicate {\it net} accretion rate, calculated subtracting the outflows (blue curve in the bottom panel) from the inflows (blue curve in the top panel). Note that the {\it net} rate can be either positive (inflow-dominated) or negative (outflow-dominated). \emph{Top panel:} The blue curve shows the inflow rate and the red curve indicates inflow-dominated times of the {\it net} accretion rate. For reference, we also show the star formation rate in the disc region as a function of time (green line). \emph{Bottom panel:} The blue curve shows the outflow rate and the red curve indicates outflow-dominated times of the {\it net} accretion rate.}
    \label{fig:acc_vs_time}
\end{figure}


\section{Results} \label{sec:results}

\subsection{Gas accretion as a function of time}

Fig.~\ref{fig:acc_vs_time} shows the gas accretion rates for Au6 as a function of time, along with the star formation rate in the disc region. In this plot, all curves indicate {\it total} accretion (i.e. disc-integrated). Blue curves show accretion rates for infall (top panel) and outflow (bottom panel) separately, which we assume to be indicated by positive and negative accretion rates, respectively.
The inflow and outflow rates have a similar time evolution, characterised by a rapid increase (in absolute values) during approximately the first $4~\mathrm{Gyr}$ of evolution  followed by a maximum accretion value at $t\sim8~\mathrm{Gyr}$, and a subsequent exponential-like decay until the present time. 

\begin{figure*}
    \centering
    \includegraphics[trim={0 0 0 0}, clip, scale=1.15]{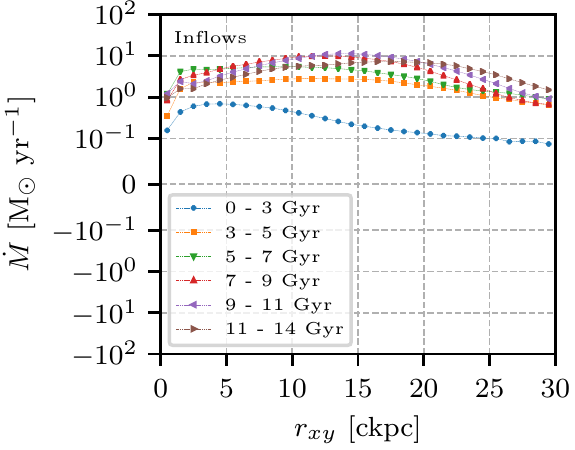}
    \includegraphics[trim={1.5cm 0 0 0}, clip, scale=1.15]{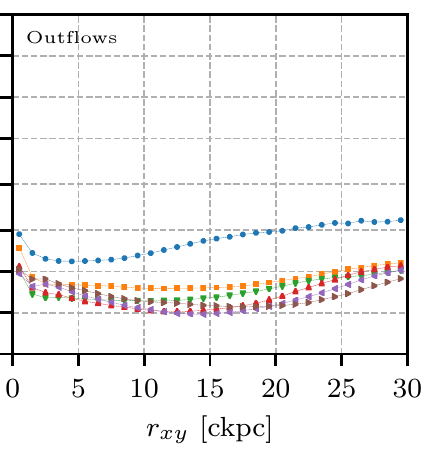} 
    \includegraphics[trim={1.5cm 0 0 0}, clip, scale=1.15]{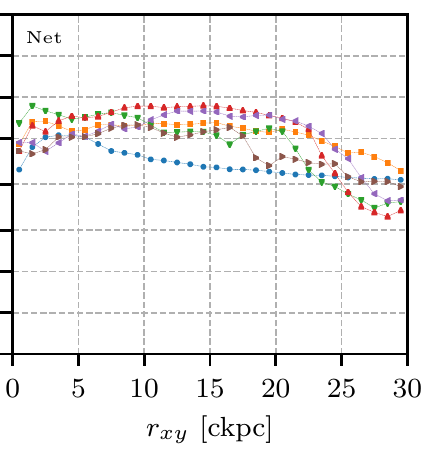}
    \caption{Accretion rates of gas as a function of comoving radius for the Au6 simulated galaxy. Each line represents the mean behaviour in a specific time interval of the galactic evolution. \emph{Left panel:} Inflow rates. \emph{Centre panel:} outflow rates, with a minus sign for a better comparison. \emph{Right panel:} Net accretion rates.}
    \label{fig:accretion_vs_radius}
\end{figure*}

The {\it net} accretion rate (infall minus outflow) follows the general trend found for inflows and outflows, being in general positive: this indicates that for most of the evolution the accretion is dominated by infall. The values obtained for the accretion rates onto the disc, with a maximum of the order of  $10^2~\mathrm{M}_\odot\,\mathrm{yr^{-1}}$,  are similar to previous results for other MW-mass galaxies \citep[e.g.][]{Nuza2019}. It is important to note, however, that approximately $15 - 25\%$ of the points are negative, indicating that outflows might also be significant.

\subsection{Gas accretion as a function of radius}

We also study the gas accretion rates as a function of radius, in order to test the usual assumption of CEMs regarding the inside-out disc formation scenario.

As mentioned above, in order to quantify the accretion as a function of radius, we divide the galactic disc in $1~\mathrm{ckpc}$-wide rings up to the maximum disc radius, and analyse the mean behaviour of the accretion rates in 6 time intervals that cover the entire evolution of the galaxy.

Fig.~\ref{fig:accretion_vs_radius} shows the gas accretion rates as a function of comoving radius for Au6.  We show trends for inflows (left), outflows (centre, shown with negative sign) and net accretion (right) separately. Each coloured line represents the mean values in a given time period of the evolution, as indicated.
This figure shows a general agreement with the inside-out disc formation scenario: at early times, accretion rates (both inflows/outflows and net) are higher for small radii while, at later times, the trend is reversed with higher accretion values for larger radii. This is more clearly seen in the left panel (inflow rate), and for radii below $20 - 25~\mathrm{ckpc}$. We stress, however, that this type of growth is not general to all galaxies and, ultimately, it is related to their particular merger and gas accretion history.

\section{Summary}

In this work, we presented an analysis of gas accretion onto the disc of a simulated galaxy from the Auriga Project. Our goal was to quantify accretion both as a function of time and galactic radius, in the disc region.

We found that inflows, outflows and net gas accretion rates follow a similar trend, with an early period 
when accretion rates rise abruptly before reaching their maximum values, and a second period characterised by a smooth decay  until the present time.

In particular, gas inflow rates, which are particularly important in CEMs, show a similar trend to those found in previous works \citep[e.g.][]{Nuza2019}, with a maximum value  of the order of $10^2~\mathrm{M}_\odot \, \mathrm{yr}^{-1}$ at $t\sim 8~\mathrm{Gyr}$ of evolution. Regarding the accretion rates as a function of radius, we found that the general behaviour follows that of the inside-out formation scenario: at early times, accretion levels are higher at small radii but at later times this trend is reversed and higher accretion levels occur at larger radii.

Our findings show that the usual assumptions of CEMs for the evolution of the Galaxy, in terms of an exponential behaviour for the time evolution, and an inside-out formation scenario,  are consistent with galaxies formed in a cosmological context. However, it is important to test these assumptions on a large number of Milky Way-mass galaxies, in order to estimate possible variations originated in differences in the formation and merger histories.

\begin{acknowledgement}
The authors acknowledge support by the Agencia Nacional de Promoción Científica y
Tecnológica (ANPCyT, PICT-201-0667).
\end{acknowledgement}


\bibliographystyle{baaa}
\small
\bibliography{main}

\end{document}